# The Gravitational Deflection of Light in F(R)-gravity

## Long Huang・Feng He・Hai Huang・Min Yao

**Abstract**   The fact that the gravitation could deflect the light trajectory has been confirmed by a large number of observation data, that is consistent with the result calculated by Einstein's gravity. F(R)-gravity is the modification of Einstein's gravity. According to the field equations obtained by the action of the f(R) form, we get a similar Schwarzschild metric. According to the condition that four-dimension momenta of the photon return to zero and that of conservation of covariant momenta, we obtain the equation of motion of the photon in a specific form of f (R)-gravity. We solve the equation to get the gravitational deflection angle of light that grazes the surface of sun and the calculation result is consistent with the experimental observation data.

## 1 Introduction

In recent decades, a large number of observation data show that the gravitational interaction causes the light bending. The observation data is consistent with the result calculated by Einstein's gravity. Einstein's gravity can reasonably explain some phenomena, such as the gravitational deflection of light, radar echo delay, the precession of Mercury perihelion, the gravitational red shift of light frequency, but can not reasonably explain cosmic acceleration (the current accelerating expansion of universe has been strongly supported by astronomical observation data; the most direct observation evidence comes from SNIa's brightness-redshift relation[1-3]; The cosmic acceleration has also been confirmed by other experimental observation

L. Huang・F. He・H. Huang
Physics College, Hunan University of Science and Technology, XiangTan 411201, Hunan Province,
People's Republic of China
F. He
e-mail: fhe@hnust.edu.cn
L. Huang
e-mail: 478196365@qq.com
M. Yao
Department of   Physics and Electronic Information Engineering, Xiangnan University, Chenzhou ,423000, China
e-mail: yaomin8162@163.com

data[4]). In recent years, according to the rotation velocity curve of the galaxy, people proposed many gravitational theories, such as dark matter model[5-6], the Modified Newtonian Dynamics (MOND) [7-8], the modified Einstein's gravity. The most simple one of the modified Einstein's gravity theories is the so-called f(R)-gravity and the scalar curvature term R in the gravitational action is replaced by a function f(R). In this paper we shall calculate the deflection angle of the light by a given metric from a specific form of f(R)-gravity and conclude finally. (In this paper we use the natural units c=1 and c is the speed of light).

## 2 $f(R) = \dfrac{2}{(1-\alpha)} \dfrac{1}{r^2} \left[ \alpha(\dfrac{r}{s})^\alpha + (2+\alpha)\lambda(\dfrac{r}{s})^2 \right]$ Gravitation Theory Model

As an alternative to Einstein-Hilbert action, we assume

$$I = I_G + I_M = -\dfrac{M_P^2}{2} \int f(R)\sqrt{-g}d^4x + \int L_M \sqrt{-g}d^4x \qquad (1)$$

where $M_P = 1/\sqrt{8\pi G}$ is Planck mass, R is the Ricci scalar and f(R) is a function of R, g is the metric determinant, $L_M$ is the Lagrangian density of the material. Variations of the action $I$ with respect to the metric tensor lead to the following field equation[9] and here we take the energy-momentum tensor is zero in vacuum.

$$R_{\mu\nu} - \dfrac{1}{2}g_{\mu\nu}\dfrac{f}{h} = (h_{;\mu\nu} - h_{;\lambda}{}^{\lambda}g_{\mu\nu})\dfrac{1}{h}, \qquad (2)$$

where $f = f(R), h = \dfrac{df}{dR}$.

For static gravitational source, the general form of the spherically symmetric metric field is

$$ds^2 = -B(r)dt^2 + A(r)dr^2 + r^2(d\theta^2 + \sin^2\theta d\varphi^2). \qquad (3)$$

Y.Sobouti gives spherically symmetric metric of a static gravitational source, whose component values are[10]

$$\dfrac{1}{A} = \dfrac{1}{1-\alpha}[1 - (\dfrac{s}{r})^{(1-\alpha/2)} + \lambda(\dfrac{r}{s})^{2(1-\alpha/2)}], \qquad (4)$$

$$B = (\dfrac{r}{s})^\alpha \dfrac{1}{A}. \qquad (5)$$

Putting the above metric into the field equation (2), we can get the specific form of f(R), namely

$$f(R) = \frac{2}{(1-\alpha)} \frac{1}{r^2} \left[ \alpha(\frac{r}{s})^\alpha + (2+\alpha)\lambda(\frac{r}{s})^2 \right], \quad (6)$$

$$R = \frac{3}{(1-\alpha)} \frac{1}{r^2} \left[ \alpha + (4-\alpha)\lambda(\frac{r}{s})^{(2-\alpha)} \right]. \quad (7)$$

Where λ is a constant. When λ=0,

$$f(R) = (3\alpha)^{\alpha/2} s^{-\alpha} R^{(1-\alpha/2)}, \quad (8)$$

where s is the large scale of the system and can be identified with the Schwarzschild radius of a center body, namely $s=2GM/c^2$, α is a small dimensionless parameter.

Y.Sobouti applies the above metric to the geodesic equation to obtain the galaxy orbital velocity formula, combines with the observation data[11], and obtains that the value of α is relative to the galaxy quality and its value is α =(3.07±0.18)×$10^{-7}$ （M/$10^{10}M_\odot$）$^{0.494}$, where $M_\odot$ is the solar mass and M is the galaxy quality.

# 3 The Gravitational Deflection Angle of the Light in $f(R)=(3\alpha)^{\alpha/2} s^{-\alpha} R^{(1-\alpha/2)}$ Gravitation Theory Model

## *3.1 Derivation of the Equation*

We choose a scalar affine parameter σ and define four-dimensional momenta of a photon as

$$p^\mu = \frac{dx^\mu}{d\sigma}. \quad (9)$$

When the photon moves along the geodesic in the gravitational field, there are still momenta $p_0$ and $p_3$ conserved [12], namely

$$r^2 \frac{d\varphi}{d\sigma} = L, \quad (10)$$

$$-B(r)\frac{dt}{d\sigma} = E. \quad (11)$$

Where $E$ is the size of the linear momentum in the distance, $L$ is the size of the angular momentum, $L/E$ is the equivalent aiming distance or collision parameter[12].

According to the condition that four-dimensional momenta return to zero, namely $g_{\mu\nu} p^\mu p^\nu = 0$, we get

$$-B(r)(\frac{dt}{d\sigma})^2 + A(r)(\frac{dr}{d\sigma})^2 + r^2(\frac{d\theta}{d\sigma})^2 + r^2 \sin^2\theta(\frac{d\varphi}{d\sigma})^2 = 0 \quad \theta = \frac{\pi}{2}. \quad (12)$$

Substitution of Eqs.(10) and (11) into Eq. (12) yields

$$-\frac{E^2}{B(r)} + A(r)(\frac{dr}{d\sigma})^2 + \frac{L^2}{r^2} = 0. \quad (13)$$

We substitute Eq.(10) into Eq. (13) to eliminate σ, which yields

$$\left[\frac{d}{d\varphi}(\frac{1}{r})\right]^2 + \frac{1}{A(r)r^2} - \frac{E^2}{L^2 A(r)B(r)} = 0. \quad (14)$$

Let λ=0 and substitution of A(r) and B(r) in Eqs.(4) and (5) into Eq.(14) yields

$$\left[\frac{d}{d\varphi}(\frac{1}{r})\right]^2 + \frac{1}{1-\alpha}\left[1 - \left(\frac{s}{r}\right)^{(1-\alpha/2)}\right]\frac{1}{r^2} - \frac{E^2}{L^2}\left(\frac{s}{r}\right)^\alpha = 0, \quad (15)$$

Let s=2GM/c² and u=GM/r, and we substitute them into Eq.(15) and take its derivative with respect to φ, which yields

$$\frac{d^2u}{d\varphi^2} + \overbrace{\frac{1}{1-\alpha}u}^{1} - \overbrace{\frac{3-\alpha/2}{1-\alpha}2^{-\alpha/2}u^{2-\alpha/2}}^{2} - \overbrace{\frac{E^2}{L^2}2^{\alpha-1}\alpha(GM)^2 u^{\alpha-1}}^{3} = 0. \quad (16)$$

This is the equation of motion of the photon in f(R) gravitation theory.

## 3.2 Simplification of the Equation

When α=0, Eq.(16) becomes

$$\frac{d^2u}{d\varphi^2} + u - 3u^2 = 0. \quad (17)$$

This is the equation of motion of photon in Einstein's gravity theory.

Now we make estimation of magnitude order to all the items in the Eq. (16) and simplify. For the sun, GM=1.5×10³m,

α =(3.07±0.18)×10⁻⁷（M/10¹⁰M☉）^0.494≈3.07×10⁻¹², where M is the quality of the solar system and approximates to solar mass.

a. $\frac{1}{1-\alpha}u$ is the first item of Eq.(16) and α≈3.07×10⁻¹² is a small quantity. So it can be approximated to u.

b. $\frac{3-\alpha/2}{1-\alpha}2^{-\alpha/2}u^{2-\alpha/2}$ is the second item of Eq.(16) and α≈3.07×10⁻¹² is a small quantity. So it can be approximated to -3u².

c. $\frac{E^2}{L^2} 2^{\alpha-1} \alpha (GM)^2 u^{\alpha-1}$ is the third item of Eq.(16), L/E is the equivalent aiming distance, namely the solar radius R (the deflection angle is the largest when the light grazes the sun edge), the magnitude order of the item is $10^{-24} u^{-1}$.

From a, b, c, the equation of motion of photon in f(R)-gravity can be simplified as

$$\frac{d^2 u}{d\varphi^2} + u = 3u^2 + 10^{-24} u^{-1}. \tag{18}$$

Compared with Eq. (17), we conclude that the main amendment in Eq. (18) is the third item, namely $10^{-24} u^{-1}$.

### 3.3 Solution of the Equation

Now we solve the equation (18):

$$\frac{d^2 u}{d\varphi^2} + u = 3u^2 + 10^{-24} u^{-1}.$$

Because u is a dimensionless and small quantity ($u_0 = \frac{GM}{R} = 2.16 \times 10^{-6}$), we omit the second-order small quantity $3u^2$ and high-order small quantity $10^{-24} u^{-1}$. The equation (18) can be changed into

$$\frac{d^2 u}{d\varphi^2} + u = 0. \tag{19}$$

The solution of Eq.(19) or the zero-order approximate solution of Eq.(18) is

$$u = u_0 \cos\varphi, \qquad u_0 = \frac{GM}{R}. \tag{20}$$

Substituting the solution (20) into the right of Eq.(18) yields

$$\frac{d^2 u}{d\varphi^2} + u = 3u_0^2 \cos^2\varphi + 10^{-24} (u_0 \cos\varphi)^{-1}. \tag{21}$$

As a linear equation, we divide Eq.(21) into two parts to solve. Let

$$u = u_1 + u_2, \tag{22}$$

where $u_1$ satisfies the equation

$$\frac{d^2 u}{d\varphi^2} + u = 3u_0^2 \cos^2\varphi, \tag{23}$$

its particular solution is

$$u_1 = u_0^2(1+\sin^2\varphi); \tag{24}$$

where $u_2$ satisfies the equation

$$\frac{d^2u}{d\varphi^2} + u = 10^{-24}(u_0\cos\varphi)^{-1}, \tag{25}$$

its particular solution is

$$u_2 = \int_{\varphi_0}^{\varphi} 10^{-24}(u_0\cos\theta)^{-1}\sin(\varphi-\theta)d\theta \qquad \varphi_0 \text{ is an arbitrary constant}$$

$$= 10^{-24}u_0^{-1}(\sin\varphi + \cos\varphi\, ln|\cos\theta|)\Big|_{\varphi_0}^{\varphi} \qquad \varphi_0 \text{ takes zero}$$

$$= 10^{-24}u_0^{-1}(\sin\varphi + \cos\varphi\, ln|\cos\varphi|). \tag{26}$$

The first-order approximate solution for the Eq.(18) is

$$u = u_0\cos\varphi + u_0^2(1+\sin^2\varphi) + 10^{-24}u_0^{-1}(\sin\varphi + \cos\varphi\, ln|\cos\varphi|). \tag{27}$$

The azimuth of the zero-order solution (20) in the distance is $\pm\frac{\pi}{2}$. The azimuth of the first-order approximate solution (27) in the distance is $\pm(\frac{\pi}{2}+\theta)$, where $\theta$ is a small quantity and satisfies the equation

$$-u_0\sin\theta + u_0^2(1+\cos^2\theta) + 10^{-24}u_0^{-1}(\cos\theta - \sin\theta\, ln|\sin\theta|) = 0. \tag{28}$$

Making Taylor expansion to $\sin\theta$ and retaining the main item, we get

$$-u_0\theta + (2-\theta^2)u_0^2 + 10^{-24}u_0^{-1}(\sqrt{1-\theta^2} - \theta\, ln\theta) = 0. \tag{29}$$

Taking $u_0 = \frac{GM}{R} = 2.16\times 10^{-6}$ and using Maple to solve the Eq.(29), we obtain

$$\theta = 4.32\times 10^{-6}. \tag{30}$$

So we eventually get the deflection angle of light

$$\Delta a = 2\theta = 1.78". \tag{31}$$

## 4 Conclusions

Since 1919, the observation team has repeatedly measured the deflection angle of the light that grazes the surface of the sun when the eclipse occurred. The measure principle is to calculate the deflection angle of the light according to the size of the positional deviation of the observed stars. So far people have measured more than four hundred stars, observation data of the deflection angle of the light range from 1.57" to 2.37", its average is 1.89", and the latest observation datum is $1.70 \pm 0.10"$. The calculation result of the deflection angle of the light by the f(R) gravitation theory is consistent with the experimental observation datum, which verifies the correctness of this f(R) gravitation theory.

## Acknowledgements

This work was supported by Hunan Provincial Natural Science Foundation of China under Grant No.11JJ6001, the Graduate Student Creative Foundation of Hunan University of Science and Technology under Grant No. S120032, National Natural Science Foundation of China under Grant No.11147011 and the Youth Scientific Research Fund of Hunan Provincial Education Department under Grant No.11B050.